\documentstyle[11pt,epsf,psfig]{article}
\def\beq{\begin{eqnarray}}
\def\eeq{\end{eqnarray}}
\def\bea{\begin{eqnarray*}}
\def\eea{\end{eqnarray*}}

\def\centeron#1#2{{\setbox0=\hbox{#1}\setbox1=\hbox{#2}\ifdim
\wd1>\wd0\kern.5\wd1\kern-.5\wd0\fi
\copy0\kern-.5\wd0\kern-.5\wd1\copy1\ifdim\wd0>\wd1
\kern.5\wd0\kern-.5\wd1\fi}}
\def\ltap{\;\centeron{\raise.35ex\hbox{$<$}}{\lower.65ex\hbox{$\sim$}}\;}
\def\gtap{\;\centeron{\raise.35ex\hbox{$>$}}{\lower.65ex\hbox{$\sim$}}\;}

\def\singleandthirdspaced{\baselineskip=\normalbaselineskip\multiply
    \baselineskip by 130\divide\baselineskip by 100}

\newcommand{\newc}{\newcommand}
\newc{\qbar}{{\overline q}}
\newc{\Kahler}{K\"ahler }
\newc{\deltaGS}{\delta_{\rm GS}}
\begin{document}
\begin{titlepage}
\begin{flushright}
{\large hep-th/yymmnnn \\ SCIPPs-2009/05\\
}
\end{flushright}

\vskip 1.2cm

\begin{center}

{\LARGE\bf Dynamics of the Peccei Quinn Scale}

\vskip 1.4cm

{\large Linda M. Carpenter, Michael Dine and Guido Festuccia}
\\
\vskip 0.4cm
{\it Santa Cruz Institute for Particle Physics and
\\ Department of Physics, University of California,
     Santa Cruz CA 95064  } \\
\vskip 4pt

\vskip 1.5cm

\begin{abstract}
Invoking the Peccei-Quinn (PQ) solution to the strong CP problem substitutes the puzzle of why $\theta_{qcd}$ is so small with the
puzzle of why the PQ symmetry is of such high quality.  Cosmological and astrophysical considerations raise further puzzles.
This paper explores this issues in several contexts:  string theory and field theory, and theories without and with low energy
supersymmetry.  Among the questions studied are whether requiring axion dark matter can account for the quality
of the PQ symmetry, to which the answer is sometimes yes.  In non-supersymmetric theories, we find $f_a = 10^{12}$ GeV
is quite plausible.  In gauge mediation, cosmological constraints on pseudomoduli place $f_a$ in this range , and require that the gravitino
mass be of order an MeV.
\end{abstract}

\end{center}

\vskip 1.0 cm

\end{titlepage}
\setcounter{footnote}{0} \setcounter{page}{2}
\setcounter{section}{0} \setcounter{subsection}{0}
\setcounter{subsubsection}{0}

\singleandthirdspaced

\section{Axions:  Their Virtues and Deficiencies}

Nuclear physics is almost indifferent to the QCD angle\cite{lorenzo}, yet for some reason $\theta_{qcd}$ is incredibly small.
In thirty years, only three persuasive solutions of this puzzle have been put forward.
\begin{enumerate}
\item  $m_u =0$:  This could result as an accident of discrete flavor symmetries\cite{banksnirseiberg}, or a result
of ``anomalous" discrete symmetries as in string theory\cite{banksdinediscrete}.
\item  Spontaneously broken CP:  Here one postulates that CP is an exact symmetry of the microscopic theory, which is spontaneously
broken.  $\theta$ is then calculable, in principle, and under certain circumstances, might
be small\cite{nelson,barr}.
In critical string theories, CP is an exact (gauge) symmetry\cite{cohenkaplancp,dinecp}, spontaneously broken at generic points in typical moduli spaces.
So this would seem a plausible framework.
\item  Axions:  in the presence of an approximate, global symmetry (Peccei-Quinn (PQ) symmetry) with a QCD anomaly, the pseudo-Goldstone boson
  which arises from symmetry breaking (the axion) adjusts to yield $\theta \approx 0$.  This raises two puzzles.  First, the symmetry
  must be {\it extremely} good if it is to solve the strong CP problem, and second, with simple, but strong, cosmological assumptions,
  the decay constant of the axion must
  be small compared to scales such as the unification scale and the Planck scale.  Critical string theories typically
  exhibit an extremely good (but approximate) global symmetry; these symmetries are exact in perturbation theory,
  broken only by non-perturbative effects\cite{wittenpq}.  Moreover, in the string theory framework, the standard
  cosmological assumptions do not hold\cite{banksdineaxion}.  So the axion solution, also, has a certain plausibility.
\end{enumerate}

Each of these solutions, however, poses problems.

\begin{enumerate}
\item  $m_u =0$.  While the work of \cite{banksnirseiberg} suggests that this possibility is often realized in models of flavor,
lattice computations appear to rule out $m_u=0$\cite{lattice}.
\item  Spontaneous CP:  To assess the plausibility of this idea, it is necessary to consider some
sort of underlying structure.  If nature is described by theories which resemble critical string theory, one needs to consider fixing of moduli.
If we take, as a model, moduli fixing in flux vacua, one often speaks of  $10^{500}$ states as arising from turning on many different fluxes.
But typically only half of the fluxes are invariant under CP.  This means that,  say  $10^{500} \rightarrow 10^{250}$ states.  So
only an {\it extremely} tiny fraction of states preserve CP.  It is not clear that these are otherwise
singled out (e.g. that most states with some other low energy feature preserve CP microscopically,
or by cosmological or anthropic considerations).  So, at least in this framework, the CP solution does not appear particularly natural.
\item  Axions:  If nature is described by a nearly critical string theory, as we will review, it is not clear, when moduli are fixed, why axions should survive to low energies.   If the
Peccei-Quinn symmetry breaking can be seen within low energy, four dimensional field theory, one cannot address the {\it quality} of this symmetry without
discussing the ultraviolet structure in which it is embedded\cite{higherdimension}.  By itself, this argument does not rule out the axion solution,
but it makes its status more uncertain.
\end{enumerate}

Given these remarks, were it not for the lattice results, it would be tempting to view the massless u quark as the most plausible
solution of the strong CP problem.  Assuming that the lattice results are verified by other groups, this will not be an option, however.
While conceptually elegant, the landscape framework, at least, calls the spontaneous CP violation solution into question.
Our goal, then, in this paper is to examine
the axion solution in various settings, and to consider carefully what is required for its successful implementation.  By successful, we mean
that the axion not be in conflict with basic facts of particle physics (the smallness of $\theta_{qcd}$) and cosmology and astrophysics (nucleosynthesis,
dark matter energy density).   By natural, we mean that there should be a sense in which the value of $f_a$ and the {\it quality} of
the axion potential should be generic.

To sharpen the notion of naturalness, we will
distinguish two theoretical frameworks for the axion:
\begin{enumerate}
\item  PQ symmetry broken by stringy or higher dimensional effects:  This case is characterized by the possibility of exponential suppression
of PQ symmetry violating operators.
\item  PQ symmetry broken in a low energy effective field theory:  Here, one needs something like a large $Z_N$ symmetry to account
for an accidental PQ symmetry\cite{dinecp}.
\end{enumerate}
Within these categories, we will consider three cases:
\begin{enumerate}
\item  No low energy supersymmetry:  Under certain circumstances, we will see
that the existence of a low value of $f_a$ is
generic.  Under the same circumstances, the requirements of a high quality axion are not as onerous as in other settings.
\item  Supersymmetry broken at an intermediate scale (``gravity mediation")
\item  Supersymmetry broken at a low scale ({\it gauge mediation})   We will touch on the main issues here, leaving a more extensive
discussion, and explicit model building, to a subsequent publication\cite{cdfu}.
\end{enumerate}

In all of these cases, one needs to ask:  what might account for the existence of a PQ symmetry?  Why should it be such a good symmetry
that it can account for the small value of $\theta_{qcd}$?
As in any discussion of naturalness, one has in mind here the notion that there is some underlying distribution of possible theories,
or states within theories,
and one views as natural choices of parameters and other features which are typical of this distribution, consistent with some
set of facts (priors)\footnote{As an example, attempts to construct measures of fine-tuning of the weak scale presuppose
a distribution of possible theories, characterized by some parameters, and ask, say, the probability of finding the observed
gauge boson masses within this distribution; theories are discarded if the probability is deemed too small.}.  One does not have to include the existence of observers among these facts, even if some of
the constraints one imposes are essential for, say, the existence of galaxies, or chemistry. One possible prior which might account
for high quality axions -- perhaps the only one -- is that the axion is a generic
way to account for the dark matter.  Imposing axion dark matter as a constraint requires that the PQ symmetry be quite good, sometimes (but
 not always) good enough to account for the smallness of $\theta_{qcd}$.
 Our remarks about the likelihood,
say, of low $f_a$, are also to be viewed in this context.

We will see that, imposing the requirement that the axion constitute the dark matter, can, in non-supersymmetric settings, potentially account for
the value of $f_a$, and possibly for the quality of the PQ symmetry.  In intermediate scale supersymmetry breaking, we are uneasy about
imposing this condition, as there are other, possibly more plausible, dark matter candidates; still, in a string/higher dimension setting,
  this condition can readily account for the small value of $\theta$ (the axion quality).  Low scale
scale breaking (gauge mediation) provides a more plausible setting for the dark
  matter condition.  Successfully implementing the axion solution in this setting places stringent requirements on the mechanism of PQ breaking,
and turns out to require a relatively high scale of supersymmetry breaking (a Goldstino decay constant $F \sim 10^{16}$ GeV$^2$), while forbidding $f_a$ much
greater than $ 10^{13}$ GeV.

The rest of this paper is organized as follows.  In section \ref{review}, we review the question of axion quality, and explain its possible connection to
dark matter.  In section \ref{stringaxions}, we discuss the question of axions in string theory or higher dimensional settings, with and without supersymmetry.
In section \ref{nonsusy}, we discuss Peccei-Quinn symmetry breaking in non-supersymmetric field theories; this is followed by a discussion
of supersymmetric field theories in section \ref{susy}.  The implications of our observations are considered in the concluding section.


\section{Axion Quality}
\label{review}

The PQ solution to the strong CP problem raises at least two serious issues.
\begin{enumerate}
\item  Astrophysics and cosmology seem to constrain the axion decay constant to a rather narrow range, $10^9$ to $10^{13} GeV $\cite{turner1,steffen}.
If the axion is to be dark matter, and if the initial axion misalignment is of order $1$,
 then $f_a \sim 10^{12}$ GeV.  Except, possibly, for the scale intermediate between the weak and the Planck
scale, this number does not correspond to other scales we suspect to be relevant to physics, such as the GUT scale or the scale
associated with neutrino masses.
\item  The PQ symmetry is a global symmetry, so it is presumably an accident.
It needs to be an extremely good symmetry if it is to solve the strong CP problem\cite{higherdimension}.
\end{enumerate}

We can easily quantify the latter problem. The contribution to the axion potential from QCD has roughly the form:
\beq
V_{qcd} \approx - m_\pi^2 f_\pi^2 \cos({a \over f_a}).
\eeq
On the other hand, the
natural value of axion potential is:
\beq
V_a =Q f_a^4 \cos({a \over f_a} - \theta_0).
\eeq
where $Q$ is a constant which we will call the {\it axion quality}.
We see that if the axion is to solve the strong CP problem, we require a suppression
of the potential below this natural value by $62$ orders of magnitude, i.e.
\beq
Q < 10^{-62} \left ({ 10^{12}  {\rm GeV} \over f_a } \right )^4.
\eeq

Things need not be as extreme as this.  In the case of low energy supersymmetry, the natural scale of the potential might be much smaller than $f_a^4$.  If the
Goldstino decay constant, $F$, is smaller than $f_a$,
 one finds that the potential is naturally suppressed by $F/f_a^2$\cite{cdfu}.

\subsection{Accounting for A Very Good Global Symmetry}

The nature of the problem is different if the Peccei-Quinn symmetry is broken at the level of four dimensional effective field theory,
or if it is broken in a higher dimensional theory or string theory.  The issues in the field theory are indicated by a simple model, with
a complex scalar field, $\phi$, on which the PQ symmetry acts as $\phi \rightarrow e^{i \alpha} \phi$, and $\langle \phi \rangle = f_a$.  An operator of the form
\beq
{\phi^{n+4} \over M_p^N}
\label{simpleaxion}
\eeq
breaks the PQ symmetry.   The ``quality factor" is given by
\beq
Q = \left ({f_a \over M_p} \right )^n.
\eeq
So, if $f_a = 10^{12}$, we require $n>10$; if $f_a = 10^{15}$, we require $n>20$!  In the optimal supersymmetric case, as explained in\cite{cdfu} we still require $n \ge 10$. The situation is somewhat better, again, if we give up the dark
matter constraint and allow for lower $f_a$.  Probably the simplest way in which one might try to account
for a suppression is through a discrete symmetry.  The symmetry would need to be quite large ($Z_{14}$ in our single field model with smaller $f_a$, $Z_{11}$
in the supersymmetric case).
We will see cases which are not so extreme shortly.  Still, such symmetries may require additional structure.

In critical string theory, the appearance of PQ symmetries in perturbation theory is a familiar phenomenon\cite{wittenpq}.    One might hope that
in a setting where moduli are fixed,
the breaking of PQ symmetries would then be governed by (powers of) a small number, such as $e^{-8 \pi^2/g^2}$, with $g^2$ some suitable
(generalized) coupling constant.  We will discuss the precise requirements, and say a little about their plausibility, later.  However, lacking anything
like a complete theory of moduli stabilization, we will not be able to make definitive statements.

\subsection{Axion as Dark Matter}

Axions have long been considered a plausible dark matter candidate. They are produced coherently in the early universe, by misalignment\footnote{We
will assume that
the reheat temperature after inflation is below $f_a$}\cite{axiondarkmatter,dfs}.  The energy density of axions is proportional to the square of the misalignment angle, $\theta_0$.
\beq
\Omega_a h^2 \approx 0.7  ~\left ( {\theta_0 \over \pi} \right)^2 \left ({f _a \over 10^{12}} \right )^{7/6}.
\label{axiondensity}
\eeq
This gives an upper bound on $f_a$, if $\theta_0 \sim \pi$, of order $10^{12}$ GeV.  If the bound is saturated, the dark matter is accounted for.
There is a lower bound coming from more conventional astrophysics of about $10^9$ GeV.

Several mechanisms have been suggested to relax the {\it upper} bound on $f_a$.  These include:
\begin{enumerate}
\item  Late decays of particles (e.g. moduli in string theory) can allow $f_a$ up to $10^{14}-10^{15}$ \cite{dfs,turner2,banksdineaxion}.
\item   Luck (or the lack of it)\cite{pi}:  If the PQ transition occurs after inflation, different regions have
different $\theta_0$.
\item  Anthropics\cite{lindeaxion,artw,freivogel}:  An elaboration on the idea above is the possibility
that anthropic considerations related to the density of dark matter select for small $\theta_0$.
Existing studies make plausible that hospitable universes lie in a narrow range of $\Omega_{dm}$, though
they hardly demonstrate this conclusively.  Note the assumption of inflation
means that, {\it if there is some peaking in the distribution}, some (anthropic) selection is inevitable.
\end{enumerate}

Having introduced in a rather non-controversial way, anthropic selection
for $\theta_0$, it is tempting to consider anthropic selection for
\begin{enumerate}
\item  The existence of axions
\item  Other parameters, such as $f_a$.
\end{enumerate}

The first point requires that, in some theoretical framework, axions be a particularly ``generic" type of dark matter.  As to the second,
in an underlying landscape, one might expect that $f_a$ varies.  This might be interesting
if requiring an axion to be the dark matter
simultaneously explains the smallness of the observed $\theta$.

\subsection{Pseudogoldstone Boson as Accidental Dark Matter}

A light pseudogoldstone boson could serve as dark matter, independent of whether it solves the strong CP problem.  For example, in the simple model
of eqn. \ref{simpleaxion}, the axion has a cosine potential and very weak coupling.  Misalignment of this field with the stationary point of its
potential will give rise to ``axion" cold dark matter.
  How light
does this axion have to be in order
to serve as dark matter?  The basic requirement is that the axion not dominate the energy density for temperatures above about $1$ eV.  If the Peccei-Quinn symmetry
is violated by some higher dimension operator, scaled by $M_p$, such as
\beq
\delta {\cal } = \left ({h \over M_p^n}  \phi^{n+4} + {\rm c.c.} \right )
\eeq
then the axion mass is:
\beq
m_a^2 = h f_a^2 \left ({f_a \over M_p}\right )^n.
\eeq
On the other hand, the initial axion energy density is of order $f_a^2/M_p^2 = 10^{-12}\left ( {f_a \over 10^{12}}\right )^2$.
So we require:
$$10^{27} \left ( {f_a \over M_p} \right )^{n + 10 \over 4} < 1$$
For $f_a = 10^{12}$, this indeed requires $n>8$, but the requirement of small enough $\theta$ means $n>10$.
So this condition is strong, but it is not, by itself, quite sufficient to account for $\theta_{qcd}$.
It is hard to assess the relative likelihood of these two cases; e.g. if the suppression
is due to a discrete symmetry, one requires a large discrete symmetry in each case, but one might worry that
a larger symmetry ($Z_{14}$) is exponentially less likely than a smaller one ($Z_{12})$\cite{dinesun}.  We will comment on this issue
further when we discuss all of this in the context of the landscape, but we will not provide a definitive answer.

Interestingly, for larger $f_a$ there is a crossover; the requirement of dark matter insures small enough $\theta$ for $f_a \sim
10^{14}$.  However, the required $n$'s are huge, more than $20$!
In string theory, within our present, limited understanding, the problem looks different, as we will discuss in the next section

\section{PQ Symmetry Broken In String Theory/Higher Dimensions}
\label{stringaxions}

As we have remarked, critical string theories seem to implement all of the known solutions to the strong CP problem.  Such theories
always have moduli, however, and the issue is whether these phenomena --  unbroken CP microscopically, axions, and discrete symmetries
(or simple accidents) which might account for a vanishing $m_u$, survive in quantum gravity theories without moduli.
At present, the only framework in which we can formulate these questions is in (hypothetical) nearly critical theories, in which
moduli are ``fixed."
It is not clear that any model of this kind exists in which systematic analysis
is possible; the most complete scenario for such moduli fixing is that of KKLT\cite{kklt}.
In any picture of moduli fixing, the problem is to understand why the non-perturbative effects
which break the PQ symmetries are extremely small (i.e. why the relevant couplings are small).
In the
simplest version of the KKLT scenario,
all moduli fixed at high scales.  The lightest is a Kahler modulus.  It's superpotential,
\beq
W = W_0 + e^{-\rho},
\eeq
explicitly breaks the would-be Peccei-Quinn symmetry.  There is a distribution of
values of $W_0$, and the small parameter arises simply because there are many possible
states, some with small $W_0$.
In the scenario, which relies heavily on approximate supersymmetry,
 $W_0 \gg m_{3/2}$, so the would-be axion lies in a massive
 chiral supermultiplet; the axion does not solve the strong CP problem\cite{bdg,donoghue}.  One might speculate that in some cases
 where there are multiple Kahler moduli, a subset would not
  appear in the superpotential, or appear suppressed by $e^{-n\rho}$, for some $n$, or suppressed
 by some other small quantity entirely.
 In the KKLT scenario, the small parameter is tied with the scale of supersymmetry breaking (up to powers of coupling constants).
\beq
e^{-\rho} = m_{3/2}^2/M_p^2
\eeq
So in order to sufficiently suppress PQ violating effects for some {\it other} modulus and solve the strong CP problem, one needs $n>3$.
We simply do not know enough about these theories to determine whether such a suppression might be generic.
As for the field theory models, we can ask whether dark matter might select for it.
The problem takes a different for depending on whether or not the low energy theory is supersymmetric.

\subsection{String Theory:  A picture without supersymmetry}

Consider, first, the possibility that there is no low energy supersymmetry, and
that there is a small parameter, $e^{-8 \pi^2 / g^2} = \epsilon$.
If $\epsilon$ is the strength of PQ breaking, and if $f_a = 10^{15}$, then we require
\beq
Q= \epsilon = 10^{-74}
\eeq
to account for $\theta$.

We can now ask whether the condition to obtain suitable dark matter is equally strong.
Again, we will take $f_a = 10^{15}$, a plausible scale for string theory; we
expect that as the axion starts to oscillate, it represents a fraction $f_a^2/M_p^2$
of the energy density.  Requiring that the axion not dominate the energy density before $T= 1$ eV,
gives $\epsilon < 10^{-78}$.
So under such circumstances, the requirement of dark matter might explain the
small value of $\theta_{qcd}$.

\subsection{String Theory With Low Energy Supersymmetry}

If supersymmetry is broken at low energies, there is at least one small parameter,
$\epsilon = {m_{3/2}^2/M_p^2}$.  Assuming that there is a PQ symmetry
violated only by terms of order $\epsilon^n$, and again taking $f_a = 10^{15}$ GeV,
the requirement that the axion yield the dark matter yields $n \ge 3$.    Again, this is enough to explain $\theta_{qcd}$.

Alternatively, there might be some other small quantity, $\epsilon^\prime \ll \epsilon$.
But we are clearly on shakier ground in imposing the requirement that the axion
is the dark matter in the framework of supersymmetry; there are other plausible candidates,
which might well arise in more generic ways (i.e. through a conserved $R$ parity).

\section{PQ Breaking Within Non-Supersymmetric Effective Field Theories}
\label{nonsusy}

In non-supersymmetric field theories, in addition to the question of axion quality, the small value of the axion decay constant is a puzzle.
We can ask, along the lines of Aguirre et al, whether not only the initial value
value of $\theta$, $\theta_0$, might be selected to account for a narrow range of dark matter densities, but
similarly $f_a$.   In a non-supersymmetric theory, we would expect small $f_a$ to be much more improbable
than small $\theta_0$; if $M_p$ is the fundamental scale, and assuming a uniform distribution of $f_a^2$, $f_a = 10^{12}$ would
be extremely improbable (unnatural);  $\theta_0 < 10^{-3}$ would seem far more reasonable.

\subsection{Dynamical Breaking of PQ Symmetry}

The situation is different the Peccei-Quinn
symmetry is {\it dynamically} broken.
Consider, for example, an  $SU(N)$ gauge theory, with a set of fields with $SU(N) \times SU(5)$ quantum numbers:
 \beq
 Q = (N, 5)~~\overline{ Q} = (\overline{ N}, \overline{ 5})~~~~ q = (N,1), ~~\overline{ q} = (\overline{ N},1).
\eeq
This model has a PQ symmetry with a QCD anomaly.  This symmetry is broken by
\beq
\langle \bar Q Q \rangle \approx \Lambda^3~~~ \langle \bar q q \rangle \approx \Lambda^3
\eeq
with $f_a \approx \Lambda$.  Now, again in a landscape context, one might expect
\beq
\Lambda = M_p e^{-{8 \pi^2 \over b_0 g^2}}.
\eeq
If  $g^2$ is uniformly distributed, small $f_a$ is favored over small $\theta_0$.

With the assumption that $\theta_0$ and $g^2$ are uniformly distributed, one can even quantify the relative likelihood of small
$f_a$ vs. small $\theta_0$.
>From equation \ref{axiondensity}
we have seen that one either needs $\theta_0 \sim 3 \times 10^{-4}$, or $f_a \approx \times 10^{12}$, or some
combination of the two.  Defining $df_a d\theta_0{\cal P}(f_a,\theta_0)$ as the fraction of the $\theta_0$, $f_a$ space allowed
by the dark matter constraint with $f_a,\theta_0$ in the volume $d\theta_0 df_a$, we have:
\beq
\int d\theta df_a P(f_a,\theta_0) \delta\left(1 - 0.7  ~\left ( {\theta_0 \over \pi} \right)^2 \left ({f _a \over 10^{12}} \right )^{7/6} \right)=1
\eeq
then
\beq
F(f_a) = \int_0^{f_a} df_a^\prime \int_0^{2 \pi} d\theta_0 P(f_a^\prime,\theta_0) \delta\left(1 - 0.7  ~\left ( {\theta_0 \over \pi} \right)^2 \left ({f _a \over 10^{12}} \right )^{7/6} \right)
\eeq
has most of its support at $f_a \sim 10^{12}$ GeV.
This is
a {\it naturalness} argument that axions might be observable in cavity experiments.

\begin{figure}[h!]
\hspace{2cm} \psfig{figure=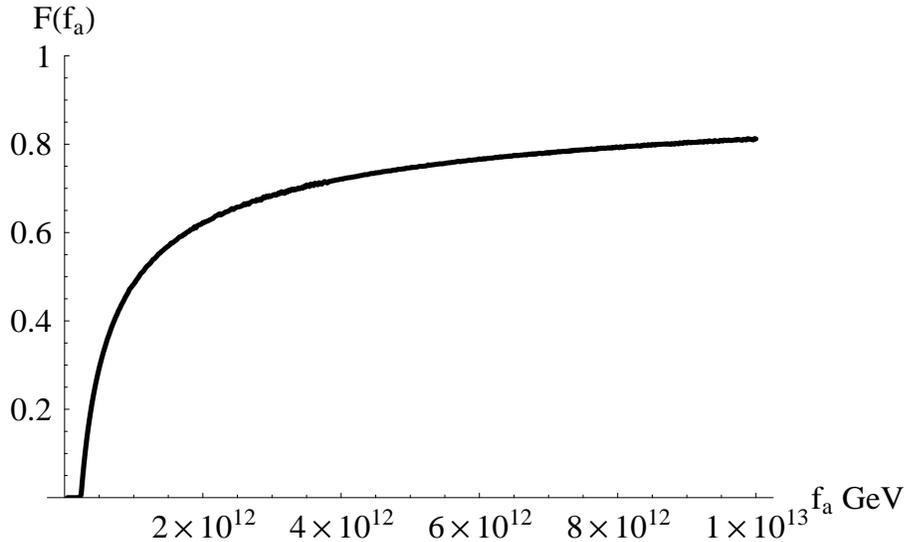,width=12cm}
\caption{The function $F(f_a)$, defined in the text.  $F= 0.8$ means that $80\%$ of the allowed range of parameters
has smaller $f_a$.}
\end{figure}

In this dynamical context, a smaller discrete symmetry might account for the quality of the PQ symmetry.  Operators allowed by the gauge symmetries, such as
\beq
\left ({1 \over M_p}\right )^{3n-4} \left ( (\bar Q Q)^n, (\bar q q)^n, {\rm etc.} \right).
\eeq
break the $U(1)$ explicitly, giving rise to an axion potential.  If $f_a = 10^{12}$ GeV, one requires
$n \ge 4$.  So a $Z_5$ symmetry might be needed to adequately suppress $\theta$.  On the other hand, a $Z_4$ symmetry is more than adequate
to yield the axion as a suitable dark matter candidate.  Arguably, the difference between $Z_4$ and $Z_5$ is not so great; moreover, the $Z_4$
symmetry is already borderline.

\section{Peccei-Quinn Breaking in Supersymmetric Field Theories}
\label{susy}

Supersymmetric theories raise new issues.  Most important, the axion decay constant is large compared to scales usually contemplated for
low energy supersymmetry breaking.  As a result, $f_a$ is determined by a (pseudo) modulus.  This particle is not necessarily the saxion;
indeed, once supersymmetry is broken, the superpartner of the axion need not be a mass eigenstate\cite{cdfu}.  The lightest modulus typically
will couple to $F_{\mu \nu}^2$ with strength comparable to that of the axion to $F \tilde F$, and similarly for the axino.

\subsection{Simple Models}

Before committing to a particular scale of supersymmetry breaking, we consider two renormalizable models which illustrate the inevitable role
of a pseudomodulus in determining the scale $f_a$.  Both possess a continuous $R$ symmetry.  Start, first, with three fields, $\chi$, $S_\pm$,
where $\chi$ possesses $R$ charge $2$ and vanishing PQ charge, while $S_\pm$ carry $R$ charge $0$ and PQ charge $\pm 1$.  The superpotential
takes the form:
\beq
W = \chi (S_+ S_- - \mu^2).
\eeq
Supersymmetry is unbroken; there is a moduli space of vacua, which we can describe by writing:
\beq
S_{\pm} = (\mu + \rho(x)) e^{\pm {\cal A}}
\eeq
Here $\rho$ and $\chi$ are massive, and $\cal A$ is the axion supermultiplet.
The axion decay constant is
\beq
f_a^2 = \mu^2 \cosh{\rm 2 Re{\cal A}}
\eeq

To break supersymmetry, we can add
fields, $X,Y$, etc, neutral under the Peccei-Quinn symmetry, along the lines of \cite{shih},which
 break supersymmetry and the $R$ symmetry.  Now if we add some ``messengers", $M, \bar M$,  carrying PQ charges and coupled
both to $S_\pm$ and $X$, one loop effects will give rise to a potential on the pseudomoduli space, ${\cal A}$\cite{cdfu}.

Alternatively, consider an O'Raifeartaigh-like model in which fields carrying $R$ charge also transform under the Peccei-Quinn symmetry.  Below, $X,Z_\pm$
carry $R$ charge $2$ and PQ charges $0, \pm 1$:
\beq
W = \lambda X(S_+ S_- - \mu^2) + m_1 Z_+ S_- + m_2 Z_- S_+.
\eeq
For large $\mu$, the PQ symmetry is broken at tree level: $\langle S_\pm \rangle \ne 0$.  Classically, there is a moduli space with
\beq
S_+ X + m_1 Z_+ = 0~~~~~ S_- X + m_2 Z_- = 0.
\eeq
On this moduli space, for large $X$, the axion decay constant is given by:
\beq
f_a^2 = \vert Z_+ \vert^2 + \vert Z_- \vert^2
\eeq
The modulus, responsible for PQ breaking, is {\it not} the saxion in this model.  Indeed, it is the partner of the $r$-axion, the  ``$r$-saxion" :
\beq
\tilde r = {\rm Re}( Z_+ + Z_-).
\eeq
While the axion and saxion arise from the orthogonal linear combination:
\beq
a = {\rm Im} (Z_+ - Z_-) ~~~  s = {\rm Re} (Z_+ + Z_-)
\eeq
which has mass of order the supersymmetry breaking scale.

At one loop, the modulus is fixed, and vanishes; the R symmetry
is unbroken  But more intricate versions of this model, following ideas of ref. \cite{shih}, yield
$R$ symmetry breaking and potentially large PQ breaking\cite{cdfu}.

\subsection{
Intermediate scale SUSY breaking}

We have earlier remarked that a scale of order $10^{12}$ GeV does not correspond, in a natural way, to other high energy scales which have been
discussed in particle physics, but this is not quite true:  in ``gravity mediation", this is the natural scale of supersymmetry breaking.  One might
then ask whether PQ breaking might be correlated with supersymmetry breaking in such a framework.  At first sight, this would
seem appealing.
This would require the axion would emerge from the hidden sector dynamics, tieing $f_a$ to the scale of supersymmetry breaking.  In that
case, the axion need not lie in an identifiable supermultiplet; there need be no saxion or axino, nor any light modulus responsible
for determining $f_a$.  The problem, however, is that in order to
generate the coupling of the axion to the Standard Model, the axion must couple to fields with mass of order this intermediate scale.
 As a result,
the sparticles of the MSSM fields would be very massive, and any connection of supersymmetry to the hierarchy problem would be lost.

The alternative, which has generally
been considered in this context, is that the axion couples only through Planck (or other large scale) suppressed operators to the hidden sector\cite{mirage}.
The axion then necessarily is accompanied by a pseudomodulus, whose value determines
$f_a$.  It is natural to call this modulus the saxion, but if there are multiple moduli, this identification may be ambiguous.
If it has a TeV scale mass, it is cosmologically problematic\cite{fischleretal}.  At about $30$ Tev, it decays before nucleosynthesis\cite{bkn}.
The situation is somewhat better if the relevant scale is lower (e.g. $M_{gut}$).  It is necessary to produce the baryon asymmetry
in these decays.
The problem of higher dimension operators is only slightly ameliorated in intermediate scale models, as the natural scale of the potential
is $M_{int}^4$.
As in the case of low energy supersymmetry in string theory, we are on shaky grounds in selecting for axions as dark matter; stable neutralinos
resulting from a conserved $R$ parity seem at least as likely to play the role of dark matter in this framework.


\subsection{
Low Scale Supersymmetry Breaking (Gauge Mediation)}

In gauge mediated models which implement a Peccei-Quinn symmetry, the scale of supersymmetry breaking is necessarily well below the
scale $f_a$.
Calling $F$ the Goldstino decay constant, one has roughly $10^5 {\rm GeV} < \sqrt{\vert F\vert} < 10^{9} {\rm GeV}$.
Suppose that the saxion couples to the messengers/susy breaking sector through Planck or Gut suppressed operators.
Even in the latter case, and for $\sqrt{\vert F \vert}= 10^9$, $m_s \sim 1$ GeV.  Its lifetime is of order
$$\Gamma \approx {m_s^3 \over M^2} \approx 10^{-32}$$
long after nucleosynthesis.  This suggests that the axion multiplet should couple directly to messengers.
Strategies for model constructions are suggested by our discussion above, and implemented in \cite{cdfu}.
This work constructs models in which either the saxion or the $r$-saxion is responsible for breaking
the Peccei-Quinn symmetry.  In these models, the saxion is not necessarily
a mass eigenstate.  What is most interesting about this analysis is that it
constrains {\it both} the scales of supersymmetry breaking and PQ breaking, as we discuss below.

\subsection{Cosmology and the scales $F$ and $f_a$ in Gauge Mediation}

Without working through models in detail, it is easy to see that cosmological considerations tend to require a large scale for the
breaking of supersymmetry -- towards the high end of what is allowed for gauge mediation.
The issue
is the mass of the light pseudomodulus, $P$.
\beq
m_P^2 = {\rm loop~factor} \times {\vert F\vert^2\over f_a^2}\sim 10^{-4} {\rm GeV^2}\left ({\sqrt{\vert F \vert} \over 10^5} \right )^4 \left ( {10^{12} \over f_a} \right )^2
\eeq
and its lifetime is of order
\beq
\Gamma = {1 \over 4\pi}{\alpha_s \over 4 \pi}^2 {m_P^3 \over f_a^2} \sim 10^{-35}~ {\rm GeV}
\left ({\sqrt{\vert F \vert} \over 10^5} \right )^6 \left ( {10^{12} \over f_a} \right )^5
\eeq
So of $f_a = 10^{12}$, we require $\sqrt{F} \sim 10^{8.5}$ if $P$ is to decay before nucleosynthesis.
In this case,
\beq
m_s \sim 10^{4} {\rm GeV}~~~ \Gamma \sim 10^{-17},
\eeq
well before nucleosynthesis.  In this case, the messengers have mass of order $f_a$.  If one does not require
that the axion be the dark matter, lower values of $F$ are possible.  These issues will be discussed in \cite{cdfu}.

\section{Conclusions}

There are a few general lessons which we take from this discussion.
\begin{enumerate}
\item  Spontaneous CP violation is not a likely solution of the strong CP problem.  In that case, assuming that the lattice result that $m_u \ne 0$ is
confirmed, the axion solution is the only viable solution.  The axion solution itself has deficiencies, and the possible mechanisms for their resolution
points to interesting physics.
\item  The existence of a PQ symmetry, of good enough quality to solve strong CP, {\it might} be correlated with the existence of dark matter
\item  In non-supersymmetric theories, low $f_a$ is natural if PQ breaking is dynamical.
\item  In string theory, the existence of suitable axions is likely correlated with the existence of a very small parameter as well as with dark matter.
\item  Implementing the PQ solution of the strong CP problem in supersymmetry points towards gauge mediation, with supersymmetry broken
at a rather high scale ($>10^8$ GeV).
 \item  Implementing the PQ solution of the strong CP problem in supersymmetry points towards gauge mediation, with supersymmetry broken
at a rather high scale ($>10^8$ GeV).  Assuming that the axion constitutes
the dark matter, it also points towards PQ breaking scales at the conventional upper limit of $10^{12}$ GeV.
t also points towards PQ breaking scales at the conventional upper limit of $10^{12}$ GeV.
\end{enumerate}

\noindent
{\bf Acknowledgements:}
We wish to thank Tom Banks, David Shih, Steve Shenker and Leonard Susskind for valuable conversations.  This work
supported in part by the U.S. Department of Energy.

\end{document}